\documentstyle[seceq]{ptptex}

\def\lesssim{\mathrel{\mathpalette\vereq<}}
\def\gtrsim{\mathrel{\mathpalette\vereq>}}
\makeatletter
\def\vereq#1#2{\lower3pt\vbox{\baselineskip1.5pt \lineskip1.5pt
\ialign{$\m@th#1\hfill##\hfil$\crcr#2\crcr\sim\crcr}}}
\notypesetlogo  

\title{
\begin{flushright}
{\small{\rm{DPNU-01-16}}}
\end{flushright}
\vspace{.5cm}
Model with Gauge Coupling Unification 
on $S^1/(Z_2 \times Z_2')$ Orbifold
}

\author{
N. {\sc Haba$^{1,2}$}\footnote{haba@eken.phys.nagoya-u.ac.jp},
T. {\sc Kondo$^2$}\footnote{tkondo@eken.phys.nagoya-u.ac.jp},
Y. {\sc Shimizu$^2$}\footnote{shimizu@eken.phys.nagoya-u.ac.jp}, \\
Tomoharu {\sc Suzuki$^2$}\footnote{tomoharu@eken.phys.nagoya-u.ac.jp}
and 
Kazumasa {\sc Ukai$^2$}\footnote{ukai@eken.phys.nagoya-u.ac.jp}
}

\inst{
$^1$Faculty of Engineering, Mie University,
Tsu Mie 514-8507, Japan\\
$^2$Department of Physics, Nagoya University,
Nagoya, 464-8602, Japan
}

\abst{
We suggest a simple grand unified theory 
 where the fifth dimensional coordinate is 
 compactified on an $S^1/(Z_2 \times Z_2')$ orbifold. 
This model contains additional
 ${\bf 10 + \overline{10}}$, (${\bf 15 + \overline{15}}$)
 and two ${\bf 24}$ Higgs multiplets in the 
 non-supersymmetric $SU(5)$ grand unified theory. 
Although this model is non-supersymmetric theory, 
 the gauge coupling unification is realized
 due to the mass splittings of Higgs multiplets
 by $S^1/(Z_2 \times Z_2')$ orbifolding. 
The constraints from the $b \rightarrow s \gamma$, 
 $\mu \rightarrow e \gamma$, 
 and electron and muon anomalous magnetic moments
 are also discussed. 
}

\begin{document}

\maketitle
\section{Introduction} \label{sec:intro}

When we consider the non-supersymmetric 
 $SU(5)$ grand unified theory (GUT), 
 one of the most serious problems is 
 that the gauge coupling constants are not
 unified at the high energy. 
However, the gauge coupling unification can be
 realized by a simple extension of the 
 standard model proposed by Murayama and Yanagida\cite{MY}, 
 where one more Higgs doublet and 
 two leptoquark scalar fields are added. 
On the other hand, 
 the five dimensional $SU(5)$ 
 GUT has been proposed where the fifth dimensional 
 coordinate is compactified on an
 $S^1/(Z_2 \times Z_2')$ 
 orbifold\cite{kawamura}\cite{kawamura2}\cite{HN}\cite{others}\cite{others2}\cite{HSSU}. 
In these models, only 
 Higgs and gauge fields (vector-like fields) can propagate in
 the bulk. 
The non-supersymmetric $SU(5)$ model 
 with ${\bf 10 + \overline{10}}$ Higgs fields
 on an
 $S^1/(Z_2 \times Z_2')$ 
 orbifold proposed by 
 Kawamura\cite{kawamura} 
 has the same field contents at the low energy 
 as the model of Ref.\cite{MY}. 
These two models with additional leptoquarks
 and a Higgs doublet at the low energy unifies the 
 three gauge couplings, but 
 the unification scale is 
 of order $10^{14}$ GeV, which
 disagrees with the present
 proton-decay experiment at Super-Kamiokande\cite{SK}.

In this paper, 
 we suggest a simple grand unified theory 
 where the fifth dimensional coordinate is 
 compactified on an $S^1/(Z_2 \times Z_2')$ 
 orbifold.
This model contains 
 not only ${\bf 10 + \overline{10}}$ 
 (we will also propose another model 
 which includes ${\bf 15 + \overline{15}}$)
 but also two ${\bf 24}$ Higgs multiplets in the 
 non-supersymmetric $SU(5)$ GUT. 
The gauge coupling unification is realized
 due to the mass splittings of Higgs multiplets 
 by $S^1/(Z_2 \times Z_2')$ orbifolding, where
 the unification scale becomes of order 
 $10^{17}$ GeV. 
The dominant proton-decay mode is 
 $p \rightarrow e^+ \pi^0$ via the exchange of the $X,Y$ gauge 
 bosons which have Kaluza-Klein masses.
However, the $X,Y$ gauge 
 bosons are too heavy for the proton decay 
 to be observed at Super-Kamiokande. 
The constraints from the $b \rightarrow s \gamma$, 
 $\mu \rightarrow e \gamma$ 
 and electron and muon anomalous magnetic moments
 will also be discussed.


\section{$SU(5)$ GUT on $S^1/(Z_2 \times Z_2')$}

We here consider the simple extension 
 of non-supersymmetric 
 $SU(5)$ GUT which contains ${\bf 5 + \overline{5}}$,
 ${\bf 10 + \overline{10}}$ (${\bf 15 + \overline{15}}$)
 and two ${\bf 24}$ Higgs multiplets. 
Only for satisfaction of the
 gauge coupling unification and 
 proton stability, this extension
 is just enough\footnote{The gauge hierarchy
 problem, which means the weak scale masses of 
 Higgs doublets 
 are not stabilized by the radiative corrections, 
 is not solved in our non-supersymmetric model.}. 
We consider the case that 
 the fifth dimensional coordinate ($= y$) 
 is compactified on an $S^1/(Z_2 \times Z_2')$ 
 orbifold, with 
 compactification radius $R$\footnote{
The five dimensional supersymmetric standard model 
 compactified on an $S^1/(Z_2 \times Z_2')$ 
 orbifold has been constructed in Refs.\cite{CSM}. 
There had been several works of the supersymmetry reduction
 by the compactification, for example, 
 in Refs.\cite{cern}\cite{LBL}. 
The extensions of the discrete symmetry and the 
 gauge symmetry are also 
 discussed in Refs.\cite{kawamoto} and
 \cite{others3}, respectively. 
}. 
We take the $Z_2$ parity operator as $P= {\rm{diag}}(1,1,1,1,1)$ and 
 the $Z_2'$ parity operator as $P'= {\rm{diag}}(-1,-1,-1,1,1)$ 
 acting on a ${\bf 5}$ representation in 
 $SU(5)$\cite{kawamura}\cite{kawamura2}. 
Under the parity transformation of $(Z_2, Z_2')$, 
 bosonic fields which can propagate in five dimensions 
 are divided into four-type of eigenstates
 according to the four-type of eigenvalues, 
 $( \pm, \pm)$\cite{kawamura}\cite{kawamura2}.

Let us show the field contents in our model. 
The contents of fermions are the same as 
 that of the ordinal $SU(5)$ GUT. 
We assume that chiral matter fermions 
 can not propagate 
 in the bulk and are localized on the 
 four dimensional wall at $y=0$ $(\pi R)$\cite{kawamura}\cite{kawamura2}\cite{HN}\cite{others}\cite{others2}\cite{HSSU}. 
The ${\bf 5} + {\bf \overline{5}}$ Higgs fields 
 as well as gauge bosons can exist in the bulk.  
Their gauge quantum numbers of the standard model, 
 eigenvalues of $Z_2 \times Z_2'$ discrete symmetries, and
 mass spectra at the tree level are shown in 
 Table I. \\
\begin{table}
\begin{center}
 \begin{tabular}[tb]{l|l|l|c|c}
rep. & 4d fields & quantum numbers & $Z_2 \times Z_2'$ parity & mass \\
 \hline\hline

 \hline
$\bf{24}$ 
 & $A_\mu^{a(2n)}$  & 
 $({\bf 8}, {\bf 1}) + ({\bf 1}, {\bf 3}) + ({\bf 1}, {\bf 1})$
 & $(+, +)$ & $\frac{2n}{R}$ \\
 & $A_\mu^{\hat{a}(2n+1)}$ & 
 $({\bf 3}, {\bf 2}) + ({\bf \overline{3}}, {\bf 2})$
 & $(+, -)$ & $\frac{2n+1}{R}$ \\
 & $A_5^{a(2n+2)}$   & 
 $({\bf 8}, {\bf 1}) + ({\bf 1}, {\bf 3}) + ({\bf 1}, {\bf 1})$
 & $(-, -)$ & $\frac{2n+2}{R}$ \\
 & $A_5^{\hat{a}(2n+1)}$   & 
 $({\bf 3}, {\bf 2}) + ({\bf \overline{3}}, {\bf 2})$
 & $(-, +)$ & $\frac{2n+1}{R}$ \\
  \hline
$\bf{5}$ 
 & $\phi_C^{(2n+1)}$  & 
 $({\bf 3}, {\bf 1})$
 & $(+, -)$ & $\frac{2n+1}{R}$ \\
 & $\phi_W^{(2n)}$ & 
 $({\bf 1}, {\bf 2})$
 & $(+, +)$ & $\frac{2n}{R}$ \\
  \hline
$\bf{\overline{5}}$ 
 & $\overline{\phi}_C^{(2n+1)}$  & 
 $({\bf \overline{3}}, {\bf 1})$
 & $(+, -)$ & $\frac{2n+1}{R}$ \\
 & $\overline{\phi}_W^{(2n)}$ & 
 $({\bf 1}, {\bf 2})$
 & $(+, +)$ & $\frac{2n}{R}$ \\

 \end{tabular}
\end{center}
\caption{The gauge quantum numbers of the standard model, 
 eigenvalues of $Z_2 \times Z_2'$ discrete symmetries, and
 mass spectra at the tree level
 of gauge bosons and 
 ${\bf{5}}+{{\bf{\overline{5}}}}$ Higgs scalars. }
\label{tb:particles1}
\end{table}

In addition to the above bosonic fields contained in
 the minimal $SU(5)$ GUT, 
 we introduce additional scalar fields. 
We propose model I and II here. 
The model I has additional ${\bf10} + {\bf\overline{10}}$
 and two ${\bf24}$ Higgs multiplets, and 
 the model II contains 
 additional ${\bf15} + {\bf\overline{15}}$ 
 and two ${\bf24}$ Higgs multiplets. 
Their gauge quantum numbers of the standard model, 
 eigenvalues of $Z_2 \times Z_2'$ discrete symmetries, and
 mass spectra at the tree level are shown in 
 Table II. 
\begin{table}
\begin{center}
 \begin{tabular}[tb]{l|l|l|c}
Rep & Quantum numbers & $Z_2 \times Z_2'$ parity & mass \\
 \hline\hline 

  \hline
$\bf{10}$ 
 &
 $({\bf 3}, {\bf 2})$
 & $(+, +)$ & $\frac{2n}{R}$ \\
 & 
 $({\bf \overline{3}}, {\bf 1}) + ({\bf 1}, {\bf 1})$
 & $(+, -)$ & $\frac{2n+1}{R}$ \\

  \hline
${\bf\overline{10}}$ 
 &
 $({\bf \overline{3}}, {\bf 2})$
 & $(+, +)$ & $\frac{2n}{R}$ \\
 & 
 $({\bf 3}, {\bf 1}) + ({\bf 1}, {\bf 1})$
 & $(+, -)$ & $\frac{2n+1}{R}$ \\

  \hline
$\bf{15}$ 
 & 
 $({\bf3}, {\bf 2})$
 & $(+, +)$ & $\frac{2n}{R}$ \\
 & 
 $({\bf 6}, {\bf 1}) + ({\bf 1}, {\bf 3})$
 & $(+, -)$ & $\frac{2n+1}{R}$ \\

  \hline
${\bf\overline{15}}$ 
 & 
 $({\bf \overline{3}}, {\bf 2})$
 & $(+, +)$ & $\frac{2n}{R}$ \\
 & 
 $({\bf \overline{6}}, {\bf 1}) + ({\bf 1}, {\bf 3})$
 & $(+, -)$ & $\frac{2n+1}{R}$ \\
 \hline
$\bf{24}$ 
 & 
 $({\bf 8}, {\bf 1}) + ({\bf 1}, {\bf 3}) + ({\bf 1}, {\bf 1})$
 & $(+, +)$ & $\frac{2n}{R}$ \\
 & 
 $({\bf 3}, {\bf 2}) + ({\bf \overline{3}}, {\bf 2})$
 & $(+, -)$ & $\frac{2n+1}{R}$ \\
 \end{tabular}
\end{center}
\caption{The gauge quantum numbers of the standard model, 
 eigenvalues of $Z_2 \times Z_2'$ discrete symmetries, and
 mass spectra at the tree level
 of additional scalars.}
\label{tb:particles2}
\end{table}
It 
 suggests that 
 only leptoquark fields 
 $({\bf 3}, {\bf 2})+({\bf \overline{3}}, {\bf 2})$, 
 which are denoted by 
 $Q$ and ${\overline Q}$, 
 have the Kaluza-Klein zero mode 
 in ${\bf10} + {\bf\overline{10}}$ 
(${\bf15} + {\bf\overline{15}}$) in the 
 model I (II).
Therefore, only ($Q+{\overline Q}$) in 
 ${\bf10} + {\bf\overline{10}}$ 
 $({\bf15} + {\bf\overline{15}})$ can survive
 in the low energy.
As for the ${\bf24}$ Higgs fields, 
 $({\bf 8}, {\bf 1})+({\bf 1}, {\bf 3})+({\bf 1}, {\bf 1})$
 components in $(SU(3)_c, SU(2)_L)$ representation
 have Kaluza-Klein zero mode and survive
 in the low energy.
The ${\bf24}$ Higgs fields do not have Yukawa couplings 
 with the matter fermions.


If there are no {\bf 24} Higgs fields, the field content 
 in the low energy is just the same 
 as the model of Ref.\cite{MY}, 
since the light $Q$ and ${\overline Q}$ 
 have Kaluza-Klein zero mode 
 while other components in ${\bf 10 +
 \overline{10}}$ 
 (${\bf15} + {\bf\overline{15}}$) 
 representation Higgs fields
 are super-heavy with Kaluza-Klein masses. 
This is the model in Ref.\cite{kawamura}. 
However, this field content suggests that
 the gauge couplings are unified 
 at $(5.0-7.8) \times 10^{14}$ GeV\cite{MY}, which disagrees
 with the current proton-decay experiments\cite{SK}. 
Thus, we introduce two more 
 additional ${\bf24}$ Higgs fields, which 
 make the unification scale higher.

We assume that  
 ${\bf10} + {\bf\overline{10}}$, 
 ${\bf15} + {\bf\overline{15}}$ and 
 two ${\bf24}$ multiplets
 have a gauge invariant 
 common mass $m_{\rm{scalar}}$, 
for simplicity. 
The magnitude of $m_{\rm{scalar}}$ must be larger
 than $1$ TeV in order not to conflict with
 the experimental data of oblique
 corrections, $S$ and $T$ parameters\cite{PT}. 

\section{Gauge Coupling Unification}

Now let us discuss the gauge coupling unification
 in our model following to the approach by 
 Hall and Nomura\cite{HN}\cite{DDG}.
We consider one loop renormalization group 
 equation for the three gauge 
 couplings. 
Three gauge couplings are unified at a unification scale $M_*$ with 
 the unified gauge coupling $g_*$. 
After compactification, 
 we need to consider threshold correction of 
 Kaluza-Klein modes\cite{HN}. 
The running of gauge coupling constants up to the one-loop 
 level is given by
\begin{eqnarray}
\label{28}
\displaystyle 
\alpha_i^{-1}(m_Z) &=& \alpha_*^{-1}(M_*)+\frac{1}{2\pi} 
\left\{ \alpha_i \ln \frac{m_{\rm{scalar}}}{m_Z}
+\beta_i \ln\frac{M_*}{m_Z}+\gamma_i\sum_{n=1}^{N_l} 
\ln \frac{M_*}{2nM_c} \right.\nonumber\\
\displaystyle    
& &\left. {} +\delta_i\sum_{n=1}^{N_l}\ln\frac{M_*}{(2n-1)M_c} \right\} ,
\end{eqnarray}
where 
 $M_c = R^{-1}$, and $N_l$ represents
 the number of the Kaluza-Klein mode which can
 propagate in the bulk, satisfying  
 $M_* = 2N_l M_c$. 
The beta functions of $\alpha_i, \beta_i$, and $\gamma_i$, 
 which are common both in model I and II,
 are
 $(\alpha_1,\alpha_2,\alpha_3) = (-1/15,-7/3,-8/3)$,
 $(\beta_1,\beta_2,\beta_3) = (64/15,-2/3,-13/3)$, and 
 $(\gamma_1,\gamma_2,\gamma_3) = (4/15,-4,-22/3)$, 
 where $\alpha_i + \beta_i$ corresponds to the beta function in 
 the standard model. 
$\gamma_i$ is induced from 
 Kaluza-Klein modes with $(+,+)$ and $(-,-)$,
 which have $2n/R$ masses. 
The difference of beta functions between 
 model I and II only appears in the values of $\delta_i$s, 
 since Kaluza-Klein modes with $(+,-)$ and $(-,+)$ 
 parity contribute to 
 $\delta_i$.
Model I and II show $(\delta_1,\delta_2,\delta_3) = (-184/15,-8,-14/3)$,
 and $(\delta_1,\delta_2,\delta_3) =
 (-164/15,-20/3,-10/3)$, respectively.

By taking the suitable linear combination of 
 gauge couplings\cite{HN} in Eq.(\ref{28}), the following relation holds in 
 both model I and II, 
\begin{eqnarray}
\displaystyle
&& 5\alpha_1^{-1}(m_Z)-3\alpha_2^{-1}
 (m_Z)-2\alpha_3^{-1}(m_Z) \nonumber \\
&& \;\;\;\;\;\;\;\;\qquad\qquad
= \frac{1}{2\pi}\left\{12\ln \frac{m_{\rm{scalar}}}{m_Z} 
+32\ln\frac{M_*}{m_Z} 
+28\sum_{n=1}^{N_l} \ln\frac{2n-1}{2n} \right\}.
\end{eqnarray}
On the other hand, 
 we can calculate the same linear combination of 
 gauge couplings in 
 the four dimensional two Higgs doublet model with 
 two leptoquarks and two $({\bf 8}, {\bf 1}) + 
 ({\bf 1}, {\bf 3}) + ({\bf 1}, {\bf 1})$ scalars,
\begin{eqnarray}
\displaystyle
5\alpha_1^{-1}(m_Z)-3\alpha_2^{-1}(m_Z)-2\alpha_3^{-1}(m_Z)
=\frac{1}{2\pi}\left\{12\ln \frac{m_{\rm{scalar}}}{m_Z} +32\ln\frac{M_U}{m_Z} \right\},\quad
\end{eqnarray}
where $M_U$ is the four dimensional unification scale in such a model. 
Thus, the model I and II relate to the 
 four dimensional model by  
%
\begin{eqnarray}
\label{MU}
\displaystyle
\ln \frac{M_{c}}{m_Z}=\ln\frac{M_U}{m_Z}+\frac{7}{8}\sum_{n=1}^{N_l}
\ln\frac{2n}{2n-1}-\ln(2N_l).
\end{eqnarray}
%
%
%
We can see the correspondence 
 between $M_c$ and $N_l$, since $M_U$ is determined 
 by the condition of gauge coupling unification.

Under the conditions that 
 $m_{\rm{scalar}}=O(1)$ TeV, $M_U\gtrsim 5 \times 10^{15}$ GeV, 
 and $\alpha_{U}^{-1}>0$, 
 the values of $M_U$ and $m_{\rm{scalar}}$ are 
 determined by the gauge coupling unification conditions 
 as 
 \begin{eqnarray}
2.5 \times 10^{17} {\rm GeV} \lesssim M_U \lesssim 5.2 \times 10^{17} {\rm GeV}, 
 \end{eqnarray}
 \begin{eqnarray}
1.2\; {\rm TeV} \lesssim m_{\rm{scalar}} \lesssim 6.9\; {\rm TeV},
 \end{eqnarray}
within the 
 experimental uncertainty of $\alpha_i$s \cite{PDG}. 
When gravity propagates in the bulk,
 the four dimensional Planck scale ($M_{\rm{pl}}$) is related to the five
dimensional Planck scale ($M_{\rm{pl}}^{(5)}$) 
as $M_{\rm{pl}}^{(5)} \simeq
({2M_{\rm{pl}}^2 M_c}/{\pi})^{1/3}$. We take $M_* < M_{\rm{pl}}^{(5)}$, which 
suggests the number of Kaluza-Klein mode as $N_l \leq 3$. 
Then we can obtain the value of $M_c$ according to
 the number of $N_l$ as follows.
\begin{eqnarray}
\label{condition}
N_l=1: \;\;\;\;\; 
&&
2.3 \times 10^{17} {\rm GeV} \lesssim M_c \lesssim 4.8 \times 10^{17}
{\rm GeV}, \nonumber \\
N_l=2: \;\;\;\;\; 
&&
1.5 \times 10^{17} {\rm GeV} \lesssim M_c \lesssim 3.1 \times 10^{17}
{\rm GeV}, \\
N_l=3: \;\;\;\;\; 
&&
1.2 \times 10^{17} {\rm GeV} \lesssim M_c \lesssim 2.4 \times 10^{17}
{\rm GeV}. \nonumber
\end{eqnarray}

The proton-decay experiment at Super-Kamiokande
 shows the lower bound of 
 $X,Y$ gauge boson mass as 
 $M_X \gtrsim 5 \times 10^{15}$ GeV\cite{SK}. 
Since the $X,Y$ bosons have the Kaluza-Klein masses
 as $M_X \sim M_c$ in this model, the above 
 ranges in Eq.(\ref{condition}) 
 satisfy the proton-decay constraint\footnote{
We approximately take $\alpha_{U} \sim 35$ in the 
 calculation of the life-time, 
 $\tau_p \sim {M_X^4}/{m_p^5/ \alpha_{U}^2}$.
}. 
Although the dominant proton-decay mode is 
 $p \rightarrow e^+ \pi^0$ via the exchange of the $X,Y$ gauge 
 in this model, 
 the $X,Y$ gauge 
 bosons are too heavy for the proton decay 
 to be observed at Super-Kamiokande.

\section{Phenomenological Constraints}

Let us show the phenomenological constraint
 in this model. 
In general, $Q$ and ${\overline Q}$ can 
 couple to the quarks and leptons as 
\begin{eqnarray}
\label{41}
\textit{L} = \lambda_{ij} \overline{\textit{d}_{iR}} 
\textit{l}_{jL} Q + \lambda_{ij}^{'}\overline{\textit{d}_{iR}} 
\textit{l}_{jL} \overline{Q}^{\dagger} + h.c.
\end{eqnarray}
where $\lambda_{ij}$ and $\lambda_{ij}'$ are
 unknown coupling constants dependent on the
 generation index $i$.
The interactions in Eq.(\ref{41}) contribute to 
 $b \to s \gamma$, $\mu \to e \gamma$ processes 
 and electron and 
 muon anomalous magnetic moments. 
As for the neutron and electron electric dipole moments,
 the contribution from above interactions are negligible, 
 since the corresponding one-loop diagrams are 
 always proportional to 
 $| \lambda_{ij}|^2$, $| \lambda_{ij}'|^2$, or
 $\lambda_{ij}^{\dagger}\lambda_{ij}'+$h.c.,which are
 real. 
We must notice here that these interactions 
 do not cause proton decay\footnote{The zero modes
 of {\bf 24} Higgs fields do not cause the 
 proton-decay, too. 
It is because {\bf 24} Higgs fields cannot interact with 
 the fermions.}.

We consider only $| \lambda_{ij}|^2$ term here,
 since one-loop diagrams which are 
 proportional to $| \lambda_{ij}'|^2$ 
 and $\lambda_{ij}^{\dagger}\lambda_{ij}'+$h.c are the 
 same order as $| \lambda_{ij}|^2$. 
At first we estimate the constraint from 
 the $b \to s \gamma$ and $\mu \to e \gamma$ 
 processes.
The effective Lagrangian for $b \to s \gamma$ and 
 $\mu \to e \gamma$ are given by
\begin{eqnarray}
\label{Br1}
{\cal L} (b \to s \gamma) = 
\frac{-e}{16\pi^2} \frac{\lambda_{si}
\lambda_{bi}^{\dagger} m_b}{m_Q^2} 
\left(
-Q_Q F \left( \frac{m_{l_i}^2}{m_Q^2} \right)
+Q_d G \left(\frac{m_{l_i}^2}{m_Q^2} \right)
\right)
\overline{s}_R\sigma_{\mu\nu}b_L F^{\mu\nu},
\qquad
\end{eqnarray}

\begin{eqnarray}
\label{Br2}
{\cal L}(\mu \to e \gamma) = 
\frac{-3e}{16\pi^2} \frac{\lambda_{ie}^{\dagger} 
\lambda_{i \mu}m_\mu}{m_Q^2} 
\left(Q_QF\left(\frac{m_{d_i}^2}{m_Q^2}\right)
+Q_d G\left(\frac{m_{d_i}^2}{m_Q^2}\right)\right)
\overline{e}_L \sigma_{\mu\nu}\mu_R F^{\mu\nu},
\qquad
\end{eqnarray}
%
%
where $m_{l_i}$ and $m_{d_i}$ denote the masses of the $i$-th generation 
charged lepton and down-type quark,respectively. 
$Q_Q$ is the electric charge of the leptoquark, 
and 
\begin{eqnarray}
F(x) &=& \frac{1}{6(1-x)^4} (1-6x+3x^2+2x^3-6x^2\ln x) ,\\
G(x) &=& \frac{1}{6(1-x)^4} (2+3x-6x^2+x^3+6x\ln x) .
\end{eqnarray}
Using Eq.(\ref{Br1}) and (\ref{Br2}), 
  we can calculate the
 branching ratios of $b\to s \gamma$ and $\mu\to e \gamma$.
For $b\to s \gamma$ we have taken into account the QCD corrections 
 at the next-leading order\cite{Kagan}.
%
%
The current experimental values are  
 $ 2.0 \times 10^{-4}\leq Br(b \to s \gamma)^{exp} 
 \leq 4.5 \times 10^{-4}$ in Ref.\cite{bsgamma} and
 $Br(\mu \to e\gamma)^{exp} \leq 1.2 \times 10^{-11}$ in Ref.\cite{PDG}.
In this model, $Br(b \to s \gamma)$ and 
 $Br(\mu \to e \gamma)$ are given by
\begin{eqnarray} 
Br(b \to s \gamma) \sim 5.0\times 10^{-9}
|\lambda_{s \tau}\lambda_{b \tau}^{\dagger}|^2,
\end{eqnarray}
\begin{eqnarray} 
Br(\mu \to e \gamma) \sim 2.1 \times 10^{-15}
|\lambda_{eb}^{\dagger} \lambda_{\mu b}|^2,
\end{eqnarray}
respectively, 
 where we take  $m_Q \sim 1.2 \:{\rm TeV}$ for the numerical estimation.
 Even if coupling $\lambda_{s \tau}$,  $\lambda_{\mu b}$ are of order one, 
the contribution to $Br(b \to s \gamma)$ and $Br(\mu \to e \gamma)$
are small compared to the experimental values. 

Next, we estimate the constraints on  
 the couplings from 
 the electron and muon anomalous magnetic moment
 experiments. 
The charged lepton anomalous magnetic moments are 
 given by
\begin{eqnarray}
\label{AMM}
\delta a_l = -\frac{3}{16\pi^2} \left(\frac{m_{l}}{m_Q}\right)^2
|\lambda_{li}|^2
\left(
 Q_Q F\left(\frac{m_{d_i}^2}{m_Q^2}\right)
+Q_d G\left(\frac{m_{d_i}^2}{m_Q^2}\right)
\right).
\end{eqnarray}
{}From Eq.(\ref {AMM}) the anomalous magnetic moments are given by
\begin{eqnarray}
\delta a_e \sim -1.3 \times 10^{-19} |\lambda_{eb}|^2,
\end{eqnarray}
\begin{eqnarray}
\delta a_{\mu} \sim -5.6 \times 10^{-15} |\lambda_{\mu b}|^2,
\end{eqnarray}
for  $m_Q \sim 1.2 \:{\rm TeV}$.
Notice that the contribution to the anomalous magnetic moments are
always negative in this model. 
Comparing with the current experimental values are 
 $\delta a_e^{exp} \simeq (31 \pm 23) 
 \times 10^{-12}$ in Refs.\cite{PDG}\cite{anomalouse}and
 $\delta a_\mu^{exp} =  (42.6 \pm 16.5) 
 \times 10^{-10}$ in Ref.\cite{anomalousmu}, 
 the magnitudes of the anomalous
 moments are small even if the coupling $\lambda_{lb}$ is of order
 one. 
Since there are still sizable theoretical and experimental
 errors, this model is not excluded by the current 
 experimental data\footnote{The current experiment of muon $g-2$ shows 
 the positive deviation from the standard model\cite{anomalousmu}}.

\section{Summary and Discussion}

We have suggested a simple grand unified theory 
 where the fifth dimensional coordinate is 
 compactified on an $S^1/(Z_2 \times Z_2')$ 
 orbifold.
This model contains 
 not only ${\bf 10 + \overline{10}}$ 
 $({\bf 15 + \overline{15}})$ 
 but also two ${\bf 24}$ Higgs multiplets in the 
 non-supersymmetric $SU(5)$ GUT. 
The gauge coupling unification is realized
 due to the mass splittings 
 by $S^1/(Z_2 \times Z_2')$ orbifolding, where
 the unification scale is of order $10^{17}$ GeV.
The dominant proton-decay mode is 
 $p \rightarrow e^+ \pi^0$ via the exchange of the $X,Y$ gauge 
 bosons which have Kaluza-Klein masses. 
However, the $X,Y$ gauge 
 bosons are too heavy for the proton decay 
 to be observed at Super-Kamiokande. 
The constraints for the $b \rightarrow s \gamma$, 
 $\mu \rightarrow e \gamma$ 
 and electron and muon anomalous magnetic moments
 have been also discussed. 
The order one couplings 
 between leptoquarks and quarks/leptons
 are consistent with the current experimental bounds.

\section*{Acknowledgment}
We would like to thank Y. Kawamura and Y. Nomura
 for useful discussions. 
Research of KU is supported in part by the Japan Society for 
 Promotion of Science under the Predoctoral Research Program. 
This work is supported in
 part by the Grant-in-Aid for Science
 Research, Ministry of Education, Science and Culture, Japan
 (No.12004276, No.12740146, No.13001292).



\begin{thebibliography}{99}


\bibitem{MY}
H.~Murayama and T.~Yanagida,
Mod.\ Phys.\ Lett.\ {\bf A7} (1992), 147. 


\bibitem{kawamura}
Y.~Kawamura,
Prog.\ Theor.\ Phys.\  {\bf 105} (2001), 691.

\bibitem{kawamura2}
Y.~Kawamura,
Prog.\ Theor.\ Phys.\  {\bf 105} (2001), 999.



\bibitem{HN}
L.~Hall and Y.~Nomura,
hep-ph/0103125.


\bibitem{others}
G.~Altarelli and F.~Feruglio,
Phys.\ Lett.\ B {\bf 511} (2001), 257.
\bibitem{others2}
A.~Hebecker and J.~March-Russell,
hep-ph/0106166.

\bibitem{HSSU}
N.~Haba, Y.~Shimizu, T.~Suzuki, and K.~Ukai,
hep-ph/0107190. 



\bibitem{SK}
M. Shiozawa {\it et al.} [Super-Kamiokande Collaboration],
Phys.\ Rev.\ Lett.\ {\bf 81} (1998), 3319.

\bibitem{CSM}
R.~Barbieri, L.~J.~Hall and Y.~Nomura,
Phys.\ Rev.\ D {\bf 63} 105007 (2001), 105007.\\
Y.~Nomura,
hep-ph/0105113.


\bibitem{cern}
I.~Antoniadis,
Phys.\ Lett.\ B {\bf 246} (1990), 377.\\
I.~Antoniadis, C.~Munoz and M.~Quiros,
Nucl.\ Phys.\ B {\bf 397} (1993), 515.\\
A.~Pomarol and M.~Quiros,
Phys.\ Lett.\ B {\bf 438} (1998), 255.\\
A.~Delgado, A.~Pomarol and M.~Quiros,
Phys.\ Rev.\ D {\bf 60} (1999), 095008.

\bibitem{LBL}
N.~Arkani-Hamed, L.~Hall, Y.~Nomura, D.~Smith and N.~Weiner,
Nucl.\ Phys.\ B {\bf 605} (2001), 81.\\
Y.~Nomura, D.~Smith and N.~Weiner,
hep-ph/0104041.\\
N.~Weiner,
hep-ph/0106021.\\
R.~Barbieri, L.~J.~Hall and Y.~Nomura,
hep-ph/0106190.\\
R.~Barbieri, L.~J.~Hall and Y.~Nomura,
hep-th/0107004.


\bibitem{kawamoto}
T.~Kawamoto and Y.~Kawamura,
hep-ph/0106163.


\bibitem{others3}
A.~Hebecker and J.~March-Russell,
hep-ph/0107039.


\bibitem{PT}
M.E.~Peskin and T.~Takeuchi,
Phys.~Rev.~Lett. {\bf 65} (1990), 964;  Phys.~Rev. {\bf D46} (1992), 381. \\
B.~Holdom and J.~Terning,
Phys.~Lett. {\bf B247} (1990), 88. \\ 
T.~Appelquist and G.~Triantaphyllou,
Phys.~Lett. {\bf B278} (1992), 345. \\ 
T.~Appelquist and J.~Terning,
Phys.~Lett. {\bf B315} (1993), 139.

\bibitem{DDG}
K.R.~Dienes, E.~Dudas and T.~Gherghetta,
Phys.~Lett. {\bf B436} (1998), 55;  Nucl.\ Phys.\ B {\bf 537} (1999), 47.
\bibitem{PDG}
Particle data group, Review of particle physics,
Eur.\ Phys.\ J.\
{\bf C15} (2000), 1. 
\bibitem{Kagan}
A.L.~Kagan and M.~Neubert,
Eur.\ Phys.\ J.{\bf C7} (1999), 5.
\bibitem{bsgamma}
S. Ahmed {\it et al.} [CLEO Collaboration],
hep-ex/9908022.

\bibitem{anomalouse}
T. Kinoshita,
Rep.\ Prog.\ Phys.\ {\bf 59} (1996), 1459.
\bibitem{anomalousmu}
H.N.~Brown {\it et al.} [Muon(g-2) Collaboration],
Phys.\ Rev.\ Lett.\ {\bf 86} (2001), 2227.  





\end{thebibliography}
\end{document}